\documentclass{elsart}
\usepackage{amsmath,amssymb}
\usepackage{natbib}
\usepackage{graphics}
\usepackage{graphicx}
\usepackage{epsfig}
\usepackage{bm}
\usepackage{amssymb}
\usepackage{lineno}

\begin{document}
\topmargin=10 true mm

\begin{frontmatter}
\title{Pattern formation of a predator-prey system
with Ivlev-type functional response}

\author[a]{Weiming Wang\corauthref{cor}},
\corauth[cor]{Corresponding author.}
\ead{weimingwang2003@163.com.}
\author[a,b]{Lei Zhang},
\author[c]{Hailing Wang}
\author[d]{Zhenqing Li}
\address[a]{Institute of Nonlinear
Analysis, School of Mathematics and Information Science, Wenzhou
University, Wenzhou, 325035 P.R. China}
\address[b]{Department of Mathematics,
North University of China, Taiyuan, Shan'xi 030051}
\address[c]{College of Computer Science and Technology, Chongqing
University of Posts and Telecommunications, Chongqing 400065}
\address[d]{Laboratory of Quantitative Vegetation Ecology,
Institute of Botany, The Chinese Academy of Sciences, Beijing
100093}

\begin{abstract}
In this paper, we investigate the emergence of a predator-prey
system with Ivlev-type functional response and reaction-diffusion.
We study how diffusion affects the stability of predator-prey
coexistence equilibrium and derive the conditions for Hopf and
Turing bifurcation in the spatial domain. Based on the bifurcation
analysis, we give the spatial pattern formation, the evolution
process of the system near the coexistence equilibrium point, via
numerical simulation. We find that pure Hopf instability leads to
the formation of spiral patterns and pure Turing instability
destroys the spiral pattern and leads to the formation of chaotic
spatial pattern. Furthermore, we perform three categories of initial
perturbations which predators are introduced in a small domain to
the coexistence equilibrium point to illustrate the emergence of
spatiotemporal patterns, we also find that in the beginning of
evolution of the spatial pattern, the special initial conditions
have an effect on the formation of spatial patterns, though the
effect is less and less with the more and more iterations. This
indicates that for prey-dependent type predator-prey model, pattern
formations do depend on the initial conditions, while for
predator-dependent type they do not. Our results show that modeling
by reaction-diffusion equations is an appropriate tool for
investigating fundamental mechanisms of complex spatiotemporal
dynamics.

\end{abstract}


\end{frontmatter}

\section{Introduction}
\label{intro}

A fundamental goal of theoretical ecology is to understand how the
interactions of individual organisms with each other and with the
environment determine the distribution of populations and the
structure of communities. Empirical evidence suggests that the
spatial scale and structure of environment can influence population
interactions\citep{Cantrell2003}. The endless array of patterns and
shapes in nature has long been a source of joy and wonder to laymen
and scientists alike. Discovering how such patterns emerge
spontaneously from an orderless and homogeneous environment has been
a challenge to researchers in the natural sciences throughout the
ages~\citep{Ben-Jacob}. The problem of pattern and scale is the
central problem in ecology, unifying population biology and
ecosystems science, and marrying basic and applied
ecology~\citep{Levin1992}. The study of spatial patterns in the
distribution of organisms is a central issue in
ecology~\citep{Levin1992,Koch1994,ClaudiaNeuhauser,
David2002,medvinsky:311,Cantrell2003, Teemu2003, Teemu2004,
Murray2004, Hawick2006, Daniel2006, Maini2006, Pearce, Baurmann,
LiuJSM, Maini2007, Shoji, Wang2007} since the pioneering work of
Alan Turing~\citep{Turing1952}. And Turing reaction-diffusion system
explains spatial patterns spontaneously forming in a perfectly
homogeneous field~\citep{Uriu2007}. The instability now identified
with Turing's name is believed to be involved in the formation of
structure in many systems of biological interest~\citep{Murray2004}.
Theoretical work has shown that spatial and temporal pattern
formation can play a very important role in ecological and
evolutionary systems. Patterns can affect, for example, stability of
ecosystems, the coexistence of species, invasion of mutants and
chaos. Moreover, the patterns themselves may interact, leading to
selection on the level of patterns, interlocking ecoevolutionary
time scales, evolutionary stagnation and
diversity~\citep{Savill1999}.

The origin of these patterns has commonly attributed to two sorts of
sources ~\citep{Levin1992}. One is a heterogeneous distribution of
abiotic factors and the other is underlying mechanisms at the level
of individuals. Patterns generated in abiotically homogeneous
environments are particularly interesting because they require an
explanation based on the individual behavior of organisms. They are
commonly called ``emergent patterns", because they emerge from
interactions in spatial scales that are much larger than the
characteristic scale of individuals~\citep{David2002}.  The
instability leads to a process that might be called differentiation
and in its simplest realization is the result of a competition
between an activator (for predator-prey system, prey) and an
inhibitor (for predator-prey system, predator) diffusing at
different rates. The results of the instability has one
characteristic property: its scale or wavelength is determined by
the concentrations of ambient species and the diffusion
coefficients, and is therefore independent of any externally imposed
length scales. In the process of morphogenesis the instability is
likely to be triggered by the increasing scale of the system: the
instability occurs once the system is large enough that it contains
several natural wavelengths of the instability~\citep{Callahan}.

The past investigations have revealed that spatial inhomogeneities
like the inhomogeneous distribution of nutrients as well as
interactions on spatial scales like migration can have an important
impact on the dynamics of ecological
populations~\citep{medvinsky:311, Murray2004}. In particular it has
been shown that spatial inhomogeneities promote the persistence of
ecological populations, play an important role in speciation and
stabilize population levels~\citep{Baurmann}. Spatial ecology today
is still dominated by theoretical investigations, and empirical
studies that explore the role of space are becoming more common due
to technological advances that allow the recording of exact spatial
locations~\citep{ClaudiaNeuhauser}.

On the other hand, as we know, our ecological environment is a huge
and highly complex system. This complexity arises in part from the
diversity of biological species, and also from the complexity of
every individual organism~\citep{Jost1998}. The relationship between
predators and their prey has long been and will continue to be one
of dominant themes in both ecology and mathematical ecology due to
its universal existence and importance~\citep{kuang98global}. A
classical predator-prey system can be written as the
form~\citep{Abram2000,David2002}:
\begin{equation}\label{A0}
\begin{array}{l}
 \dot{u}=uf(u)-vg(u,v),\qquad
 \dot{v}=h[g(u,v),v] v.
\end{array}
\end{equation}
where $u$ and $v$ are prey and predator density, respectively,
$f(u)$ the prey growth rate, $g(u, v)$ the functional response, the
prey consumption rate by an average single predator, which obviously
increases with the prey consumption rate, and can be influenced by
the predator density, $h[g(u, v), v]$ the per capita growth rate of
predators (also known as the ``predator numerical response"). The
most widely accepted assumption for the numerical response is the
linear one~\citep{Arditi1989,David2002}:
$$
h[g(u, v), v]=\varepsilon g(u, v)-\beta,
$$
where $\beta$ is a per capita predator death rate and $\varepsilon$
the conversion efficiency of food into offsprings.

In population dynamics, a functional response $g(u,v)$ of the
predator to the prey density refers to the change in the density of
prey attached per unit time per predator as the prey density changes
~\citep{Ruan:1445}. In general, functional response can be
classified as (i) prey dependent, when prey density alone determines
the response, $g(u,v)=p(N)$; (ii) predator dependent, when both
predator and prey populations affect the response. Particularly,
when $g(u,v)=p(\frac{u}{v})$, we call model~\eqref{A0} strictly
ratio dependent; and (iii) multi-species dependent, when species
other than the focal predator and its prey species influence the
functional response~\citep{Abram2000}.

There have been several famous functional response types: Holling
type I--III~\citep{Holling1, Holling2}; Hassell-Varley
type~\citep{Hassell}; Beddington-DeAngelis type by
Beddington~\citep{Beddington} and DeAngelis, Goldstein and
Neill~\citep{DeAngelis} independently; the Crowley-Martin
type~\citep{Crowley}; and the recent well-known ratio-dependence
type by Arditi and Ginzburg~\citep{Arditi1989} later studied by
Kuang and Beretta~\citep{kuang98global}. Of them, the Holling type
I---III was labeled ``prey-dependent" and the other types that
consider the interference among predators were labeled
``predator-dependent"~\citep{Arditi1989}.

Besides Holling type I--III, there is another important
prey-dependent functional response--Ivlev-type, originally due to
Ivlev~\citep{Ivlev}:
\begin{eqnarray}\label{A1}
 g(u,v)=1-e^{-\gamma u}.
\end{eqnarray}
and the corresponding Ivlev-type predator-prey model takes the form:
\begin{eqnarray}\label{eq:1}
 \dot{u}=u(1-u)-v(1-e^{-\gamma u}), \qquad
 \dot{v}=\varepsilon v(1-e^{-\gamma u})-\beta v.
\end{eqnarray}
where $u$ and $v$ represent population density of prey and predator
at time $t$, respectively, $\varepsilon,\, \beta,\, \gamma$ are
positive constants, $\varepsilon$ the conversion rate of prey
captured by predator, $\beta$ the deathrate of predator, and
$\gamma$ the efficiency of predator capture of prey. From an
ecological viewpoint, the conditions $\dot{u}>0$ and $\dot{v}>0$
must hold. From the second equation of~\eqref{eq:1}, we know
$\varepsilon>\beta$.

In this paper, we mainly focus on the following spatial Ivlev-type
predator-prey model with reaction diffusion:
\begin{eqnarray}\label{eq:2}
\begin{array}{l}
 \dot{u}=\overbrace{u(1-u)}^{\text{growth due to prey}}-\overbrace{v(1-e^{-\gamma u})}
 ^{\text{mortality due to prey}}+\overbrace{d_1\nabla^2u}^{\text{random
 motility}}\equiv f(u,v)+d_1\nabla^2u, \\[4pt]
 \dot{v}=\underbrace{\alpha\beta v(1-e^{-\gamma u})}_{\text{growth due to predator}}
 -\underbrace{\beta v}_{\text{mortality due to predator}}+\underbrace{d_2\nabla^2v}
 _{\text{random motility}}\\[4pt]
 \,\,\,\,\equiv g(u,v)+d_2\nabla^2v.
\end{array}
\end{eqnarray}
where $\alpha=\frac{\varepsilon}{\beta}>1$, and $d_1, d_2$ are the
diffusion coefficients of prey and predator, respectively,
$\nabla^2=\frac{\partial}{\partial x^2}+\frac{\partial}{\partial
y^2}$ is the usual Laplacian operator in two-dimensional space.

Both ecologists and mathematicians are interested in the Ivlev-type
predator-prey model and much progress has been seen in the study of
model~\eqref{eq:1}~\citep{May1981, Metz1986, Kooij,Sugie, Tian2006,
Wang,Preedy2006} and model~\eqref{eq:2}
~\citep{Sherratt1995,Sherratt1997,Sherratt2000,Pearce,
Garvie,Preedy2006,Uriu2007}. The results indicate that the
Ivlev-type predator-prey model~\eqref{eq:1} and \eqref{eq:2} have
widely applicabilities in ecology, such as dynamics in predator-prey
system~\citep{May1981, Metz1986,Sherratt1995,
Sherratt1997,Tian2006}, host-parasitoid system~\citep{Pearce,
Preedy2006}, fish skin pattern~\citep{Uriu2007}, and so on.

Of them, Sherratt and co-workers had studied the dynamics of
oscillations and chaos behind predator-prey
invasion~\citep{Sherratt1995, Sherratt1997}. Especially, in
reference~\citep{Sherratt1997}, the authors performed a large number
of numerical simulations of the invasion of prey by predator with
four categories models, involving one- and two-dimensional
reaction-diffusion model~\eqref{eq:2}. For model~\eqref{eq:2}, they
used a large spatial domain, with the system initially in the
prey-only steady state, except for a small region in the center of
the domain, where a small density of predators was introduced. They
stopped their simulations before the invading wave reached the end
of the domain, so that the results were not sensitive to the
boundary conditions, which could be either zero flux, periodic, or
with population levels fixed at the prey-only steady state. They
also discussed the way in which the populations evolved after the
invasion had reached the edge of the domain. Furthermore, the
authors performed a number of one-dimensional spatial patterns with
the model~\eqref{eq:2} initially in the prey-only steady state and
two-dimensional spatial patterns with two categories of initial
perturbation. One is the introduction of predators are in a small,
localized region of the domain, which is otherwise in the prey-only
steady state; the other is the introduction of predators along a
line running parallel to one edge of the (rectangular) spatial
domain. Based on these results, the authors indicated that for
model~\eqref{eq:2}, the behavior behind the invasive front of
predators consists of either irregular spatiotemporal oscillations,
or periodic waves in population density.

But to our knowledge, for model~\eqref{eq:2}, the research on
symbolic conditions of Hopf and Turing bifurcation, the evolution
process of the spatial pattern formation, the mechanism of pattern
formation emergence, especially the influences of the specific
choice of the initial conditions to the pattern formation, seems
rare.

The paper is organized as follows: In Section 2, we employ the
method of stability analysis to derive the symbolic conditions for
Hopf and Turing bifurcation in the spatial domain. Based on these
conditions we locate the Hopf and Turing bifurcation within the
generalized parameter domain in $\gamma-d_1$ bifurcation diagram. In
Section 3, we give the spatial pattern formation, the evolution
process of the system near the coexistence equilibrium point, via
numerical simulation. For the sake of learning the influences of the
initial conditions to pattern formation, we perform three categories
of initial perturbations which predators are introduced in a small,
localized region of the circle, line and pitchfork domain to the
coexistence equilibrium point to illustrate the emergence of
spatiotemporal patterns. Then, in the last section, we give some
discussions and remarks.

\section{Stability and Bifurcation analysis}

The non-spatial model~\eqref{eq:1} has at most three equilibria
(stationary states), which correspond to spatially homogeneous
equilibria of the model~\eqref{eq:2}, in the positive quadrant:

(i) (0, 0) (total extinct) is a saddle point;

(ii) $(1,0)$ (extinct of the predator, or prey-only) is a saddle
when $\gamma>-\ln\frac{\alpha-1}{\alpha}$, or stable node when
$\gamma<-\ln\frac{\alpha-1}{\alpha}$, or saddle-node when
$\gamma=-\ln\frac{\alpha-1}{\alpha}$;

(iii) a nontrivial stationary state $(u^*, v^*)$ (coexistence of
prey and predator), where
\begin{equation}\label{eq:3}
\begin{array}{l}
u^*=-\frac{1}{\gamma}\ln\Bigl(\frac{\alpha-1}{\alpha}\Bigr),\qquad
v^*=-\frac{\alpha}{{\gamma}^{2}} \ln\Bigl({\frac
{\alpha-1}{\alpha}}\Bigr)\Bigl(\gamma+\ln( {\frac
{\alpha-1}{\alpha}})\Bigr) =\frac{u^*(1-u^*)}{1-e^{-\gamma u^*}}.
\end{array}\end{equation} with $\alpha>1,\,\,\gamma>-\ln\Bigl(\frac
{\alpha-1}{\alpha}\Bigr)$.

In this paper, we mainly focus on the dynamics of nontrivial
stationary state $(u^*, v^*)$. For cyclical populations, this
coexistence state will also be unstable and will lie inside a stable
limit cycle in the kinetic phase plane.

To perform a linear stability analysis, we linearize the dynamic
system~\eqref{eq:2} around the spatially homogenous fixed
point~\eqref{eq:3} for small space- and time-dependent fluctuations
and expand them in Fourier space
$$
\begin{array}{l}
u(\vec{x},t)\sim u^*e^{\lambda t}e^{i\vec{k}\cdot\vec{x}},\quad
v(\vec{x},t)\sim v^*e^{\lambda t}e^{i\vec{k}\cdot\vec{x}}.
\end{array}
$$

Then, in the linearized version of model~\eqref{eq:2}, yielding a
dispersion relation from which one can choose parameters to allow
only some of the modes with $\text{Re}(\lambda)>0$ to grow in time.
The dispersion relation $\lambda(k)$ relating the temporal growth
rate to the spatial wave number $k$ can be found from the
characteristic polynomial of the original problem~\eqref{eq:2}:
\begin{equation}\label{eq:4}
\lambda^2-\text{tr}_k\lambda+\Delta_k=0,
\end{equation}
where $$\begin{array}{l}
\text{tr}_k=f_u+g_v-k^2(d_1+d_2)\equiv\text{tr}_0-k^2(d_1+d_2),\\[8pt]
\Delta_k=f_ug_v-f_vg_u-k^2(f_ud_2+g_vd_1)+k^4d_1d_2\equiv\Delta_0-k^2(f_nd_2+g_pd_1)+k^4d_1d_2,
\end{array}
$$
here $\text{tr}_0=f_u+g_v$, $\Delta_0=f_ug_v-f_vg_u$, and the
elements $f_u, f_v, g_u, g_v$ are the partial derivatives of the
reaction kinetics $f(u,v)$ and $g(u,v)$ denoted by~\eqref{eq:2},
evaluated at the stationary state $(u^*, v^*)$.

The reaction-diffusion systems have led to the characterization of
two basic types of symmetry-breaking bifurcations---Hopf and Turing
bifurcation, responsible for the emergence of spatiotemporal
patterns. See, for details, references~\citep{yang:7259, Wang2007}.

The Hopf bifurcation occurs when $\text{Im}(\lambda(k))\neq 0$ and
$\text{Re}(\lambda(k))=0$ at $k=0$, and the critical value of Hopf
bifurcation parameter $\gamma$ equals
\begin{equation}\label{eq:5}
\gamma_H=\frac{A(2+A)}{(1-\alpha)(1+A)},
\end{equation}
where $A=(\alpha-1)\ln\frac{\alpha-1}{\alpha}$.

At the Hopf bifurcation threshold, the temporal symmetry of the
system is broken and gives rise to uniform oscillations in space and
periodic oscillations in time with the frequency
\begin{equation}\label{eq:6}
\omega_H=\text{Im}(\lambda(k))=\sqrt{\Delta_0}=\frac{\beta
A\Bigl(\gamma (\alpha-1)+A\Bigr)}{\gamma(\alpha-1)},
\end{equation}
and the corresponding wavelength is
\begin{equation}\label{eq:7}
\lambda_H=\frac{2\pi}{\omega_H}=\frac{2\gamma(\alpha-1)\pi}{\beta
A(\gamma (\alpha-1)+A)}.
\end{equation}

The Turing instability is dependent not upon the geometry of the
system but only upon the reaction rates and diffusion. And it cannot
be expected when the diffusion term is absent and it can occur only
when the activator (for predator-prey system, prey) diffuses more
slowly than the inhibitor (for predator-prey system, predator).
Mathematically speaking, as $d_1\ll d_2$, the Turing bifurcation
occurs when $\text{Im}(\lambda(k))=0$ and $\text{Re}(\lambda(k))=0$
at $k=k_T\neq 0$, $k_T$ is called the wavenumber. The critical value
of bifurcation parameter $\gamma$ equals
\begin{equation}\label{eq:8}
\begin{array}{l}
\gamma_T=\frac{A(\beta A+2d_2k_T^2+d_2Ak_T^2)}{(1-a)(\beta
A+d_2k_T^2+d_2Ak_T^2-d_1d_2k_T^4)},
\end{array}
\end{equation}
where $$k_T^2=\sqrt{\frac{\Delta_0}{d_1d_2}}.$$ And at the Turing
threshold, the spatial symmetry of the system is broken and the
patterns are stationary in time and oscillatory in space with the
wavelength
\begin{equation}\label{eq:9}
\lambda_T=\frac{2\pi}{k_T}=\frac{2\pi\sqrt[4]{d_1d_2\gamma^2(\alpha-1)^2}}{\sqrt{\beta
A(\gamma(a-1)+A)}}.
\end{equation}

\begin{figure*}[htp]
\includegraphics[width=14cm]{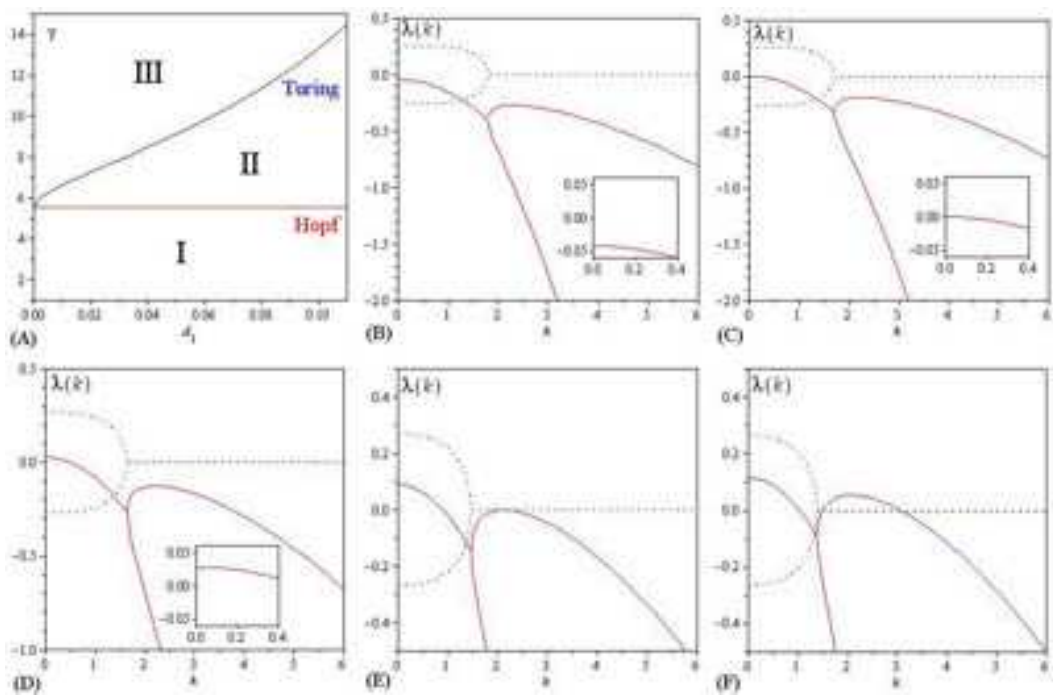}
\caption{\label{fig1} (A) $\gamma-d_1$ Bifurcation diagram for the
system~\eqref{eq:2} with $\alpha=1.1$, $\beta=0.5$, $d_2=0.2$. Hopf
and Turing bifurcation lines separate the parameter space into three
domains. The other parameters in figures (B)--(F): $d_1$=0.02, and
(B) $\gamma=5.0$; (C) $\gamma=5.552147102$, the critical value of
$\gamma_H$; (D) $\gamma=6.0$; (E) $\gamma=7.265163898$, the critical
value of $\gamma_T$; (F) $\gamma=8.0$. The real and the imaginary
parts of $\lambda(k)$ are shown by solid lines and dotted lines,
respectively.}
\end{figure*}

Linear stability analysis of model~\eqref{eq:2} yields the
bifurcation diagram, the relation between $\gamma$ and $d_1$, is
shown in figure~\ref{fig1}(A). In this case, the Hopf bifurcation
line and the Turing bifurcation line separate the parametric space
into three distinct domains. In domain I, located below all Hopf and
Turing bifurcation lines, the steady state is the only stable
solution of the system. Domain II is the region of pure Hopf
instability, and in domain III, located above all two bifurcation
lines, both Hopf and Turing instability occur.

From the definition of Hopf and Turing bifurcation, we know that the
relation between the real, the imaginary parts of the eigenvalue
$\lambda(k)$ determine the bifurcation type. The relation between
$\text{Re}(\lambda(k))$, $\text{Im}(\lambda(k))$ and $k$ are shown
in figure~\ref{fig1}(B)--(F). Figure~\ref{fig1}(B) illustrate the
case of parameter locate in domain I in figure~\ref{fig1}(A),
$\gamma=5.0$, one can see that $\text{Re}(\lambda(k))<0$ and
$\text{Im}(\lambda(k))\neq0$ at $k=0$. Figure~\ref{fig1}(C),
$\gamma=5.552147102\equiv \gamma_H$, the critical value of Hopf
bifurcation, in this case, $\text{Re}(\lambda(k))=0$ at $k=0$ while
$\text{Im}(\lambda(k))\neq0$. In figure~\ref{fig1}(D), $\gamma=6.0$,
the parameter locate in domain II, the pure Hopf instability occurs,
one can see that at $k=0$, $\text{Re}(\lambda(k))>0$,
$\text{Im}(\lambda(k))\neq0$. Figure~\ref{fig1}(E),
$\gamma=7.265163898\equiv \gamma_T$, the critical value of Turing
bifurcation, at $k=k_T=2.116874108$,
$\text{Re}(\lambda(k))=\text{Im}(\lambda(k))=0$. When $\gamma=8.0$,
parameter locate in domain III, figure~\ref{fig1}(F) indicate that
at $k=0$, $\text{Re}(\lambda(k))>0$, $\text{Im}(\lambda(k))\neq0$.

\section{Pattern formation analysis}

\begin{figure*}[htp]
\includegraphics[width=14cm]{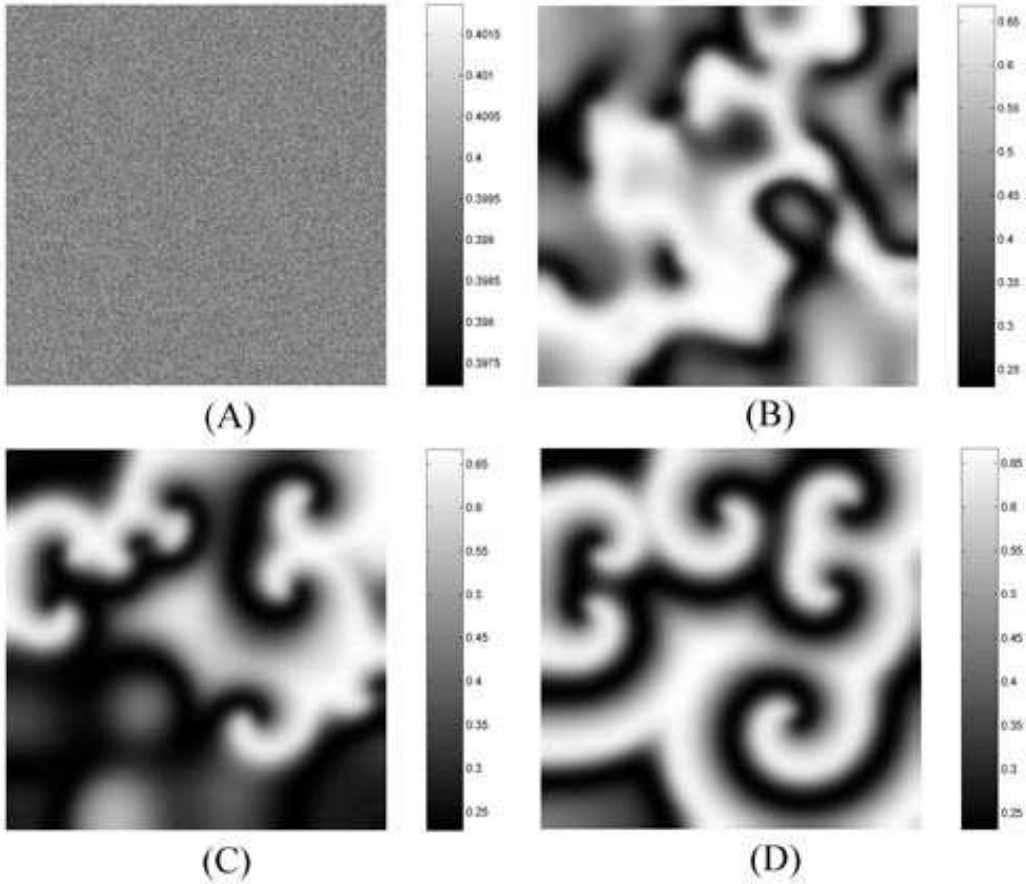}
\caption{\label{fig2} Grey-scaled snapshots of contour pictures of
the time evolution of the prey of model~\eqref{eq:2} with
$\gamma=6.0>\gamma_{H}$. (A) 0 iteration; (B) 10000 iteration; (C)
20000 iteration; (D) 50000 iteration.}
\end{figure*}

\begin{figure*}[htp]
\includegraphics[width=14cm]{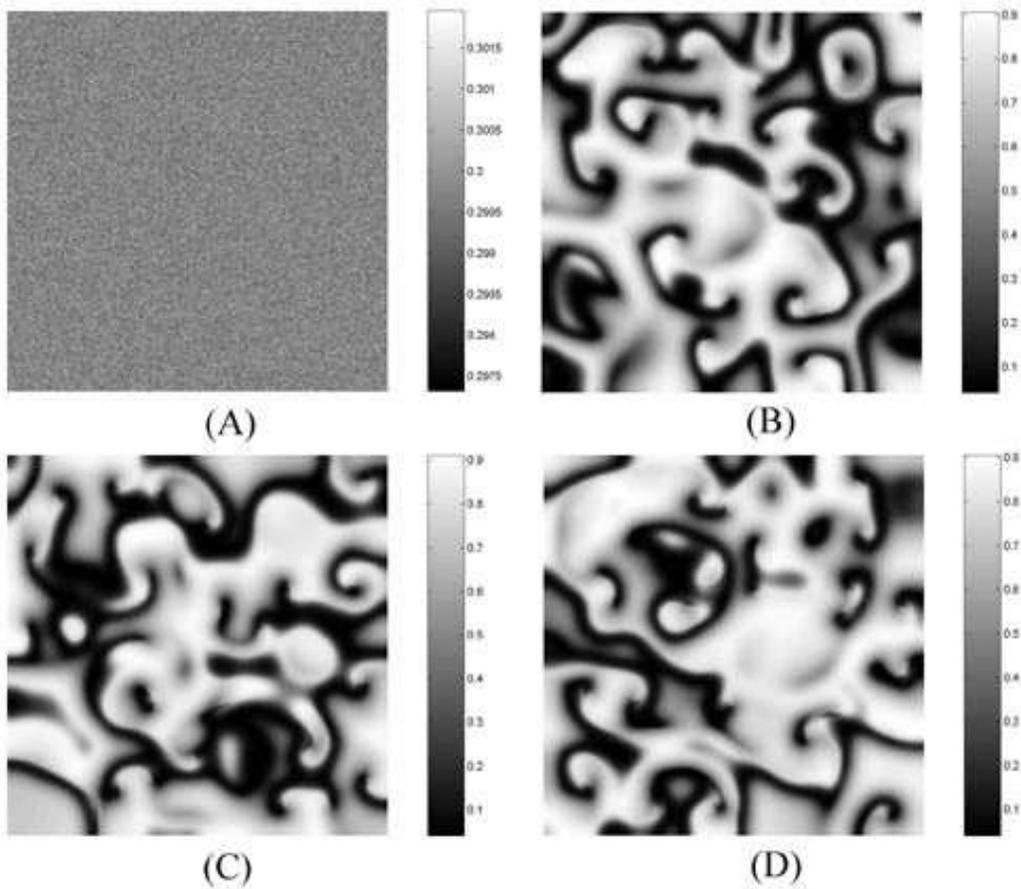}
\caption{\label{fig3} Grey-scaled snapshots of contour pictures of
the time evolution of the prey of model~\eqref{eq:2} with
$\gamma=8.0>\gamma_{T}$. (A) 0 iteration; (B) 10000 iteration; (C)
20000 iteration; (D) 50000 iteration.}
\end{figure*}

\begin{figure*}[htp]
\includegraphics[width=14cm]{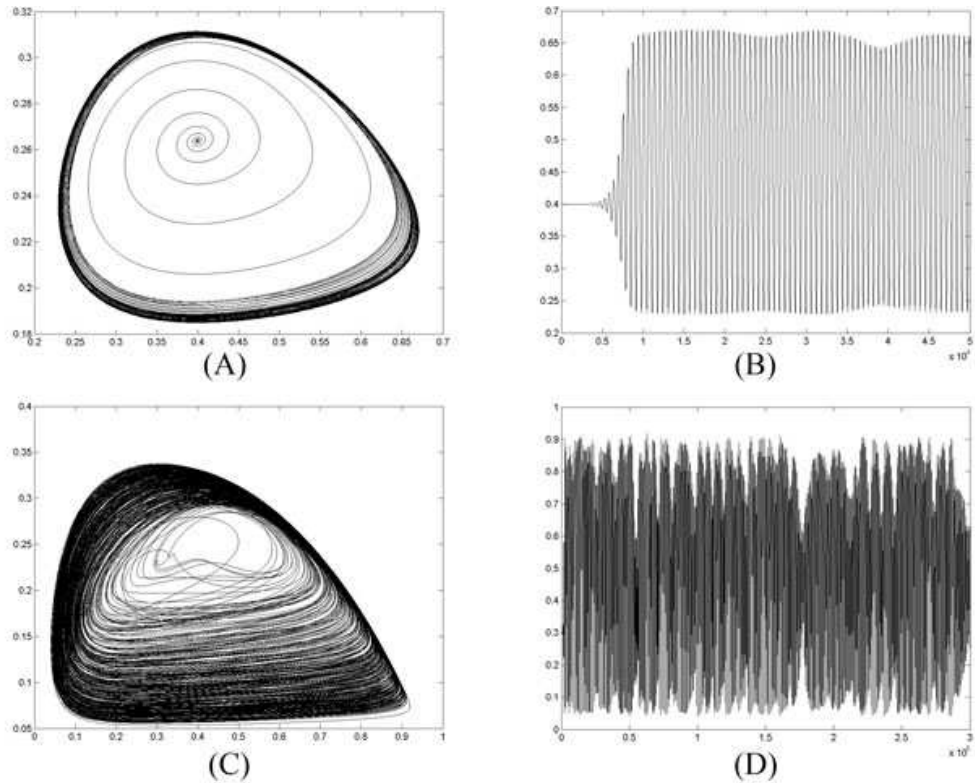}
\caption{\label{fig4} Dynamical behavior of model~\eqref{eq:2}.
Phase planes of the model at a fixed point inside the domain
occupied by irregular spatiotemporal oscillations with $\gamma=6$
(A) and $\gamma=8$ (C), respectively. Time-series plots with
$\gamma=6$ (B) and $\gamma=8$ (D), respectively.}
\end{figure*}

In this section, we have performed extensive numerical simulations
of the spatially extended model~\eqref{eq:2} in two-dimensional
space, and the qualitative results are shown here.
Model~\eqref{eq:2} is posed on a given domain $\Omega=400\times
400$, with smooth boundary $\partial\Omega$. Zero-flux Neumann
boundary conditions are imposed on $\partial\Omega$ to close the
system. Model~\eqref{eq:2} is solved numerically with $\alpha=1.1$,
$\beta=0.5$, $d_1=0.02$, $d_2=0.2$ in two-dimensional space using a
finite difference approximation for the spatial derivatives and an
explicit Euler method for the time integration~\citep{Garvie} with a
time stepsize of $\Delta t={1}/{3}$ and space stepsize $\Delta
h={1}/{24}$. We start with the unstable uniform solution $(u^*,
v^*)$ with small random perturbation superimposed. Thus the initial
profiles of $u$ and $v$ are completely random without any spatial
correlation. And we perform a series of of two-dimensional
simulations (figures~\ref{fig2}, ~\ref{fig3} and~\ref{fig4}), in
each, the initial condition was always a small amplitude random
perturbation $(\pm 5\times 10^{-4})$ around the steady state $(u^*,
v^*)$, and patterns developed spontaneously.

In the numerical simulations, different types of dynamics are
observed and we have found that the distributions of predator and
prey are always of the same type. Consequently, we can restrict our
analysis of pattern formation to one distribution. In this section,
we show the distribution of prey, for instance.

From the analysis in section 2 and the bifurcation diagram
(figure~\ref{fig1}(A)), the results of numerical simulations show
that when parameters $\alpha$, $\beta$, $d_2$ are determined, the
type of the system dynamics is determined by the values of $\gamma$
and $d_1$. And for different sets of parameters, the features of the
spatial patterns become essentially different when $\gamma$ exceeds
the critical value $\gamma_H$ and $\gamma_T$ respectively, which
depend on $d_1$.

Figure~\ref{fig2} shows the evolution of the spatial pattern of prey
at $0, 10000, 20000$ and $50000$ iterations with $\gamma=6.0$, more
than the Hopf bifurcation threshold $\gamma_{H}=5.552147102$ and
less than the Turing bifurcation threshold $\gamma_T=7.265163898$.
In this case, one can see that for model~\eqref{eq:2}, the random
initial distribution around the steady state $(u^*, v^*)=(0.39965,
0.26392)$ leads to the formation of the spiral wave pattern in the
domain (figure~\ref{fig2}(D)). In other words, in this situation,
spatially uniform steady-state predator-prey coexistence is no
longer. Small random fluctuations will be strongly amplified by
diffusion, leading to nonuniform population distributions. From the
analysis in section 2, we find with these parameters in domain II,
the spiral pattern arises from Hopf instability.

When $\gamma=8.0>\gamma_T=7.265163898$, in this case, parameters in
domain III (figure~\ref{fig1}(A)), both Hopf and Turing
instabilities occur. The nontrivial stationary state is $(u^*,
v^*)=(0.29974, 0.23088)$. As an example, the formation of a regular
macroscopic two-dimensional spatial pattern, the chaotic spiral
pattern, is shown in figure~\ref{fig3}.

Comparing this situation (figure~\ref{fig3}) with the one above
(figure~\ref{fig2}), we can see that the formation of chaotic spiral
patterns (figure~\ref{fig3}(C, D)) are caused by Turing instability.

For the sake of learning the dynamics of model~\eqref{eq:2} further,
we illustrate the phase portraits and time-series plots in
figure~\ref{fig4}. From figure~\ref{fig4}(A), for $\gamma=6.0$, one
can see that a quasi limit cycle arises, which is caused by the Hopf
bifurcation. Furthermore, we can calculate that the frequency of
periodic oscillations in time (figure~\ref{fig4}(B)) is
$\omega=27.5899$, and corresponding wavelength $\lambda=0.2277$. And
from~\eqref{eq:6} and~\eqref{eq:7}, we know that at the critical
value of Hopf bifurcation $\gamma_H$, the frequency of the periodic
oscillations in time is $\omega_H=24.0748$, the corresponding
wavelength $\lambda_H=0.2610$. When $\gamma=8.0$, the dynamical
behavior are shown in figures~\ref{fig4}(C) and (D). From
figure~\ref{fig4}(C), one can see that there exhibits a ``local"
phase plane of the system invaded by the irregular spatiotemporal
oscillations. Instead of the limit cycle in the case above
(figure~\ref{fig4}(A)), as happens in the case of smooth pattern
formation, the trajectory now fills nearly the whole domain inside
the limit cycle. This regime of the system dynamics corresponds to
spatiotemporal chaos. And the spatial symmetry of model~\eqref{eq:2}
is broken and the patterns are oscillatory in space with the
wavelength $\lambda=0.1629$ while at the critical value of Turing
bifurcation, $\lambda_T=0.2834$.

The above results are obtained from the initial conditions which was
always a small amplitude random perturbation around the steady state
$(u^*, v^*)$. In references~\citep{medvinsky:311, Sherratt1997,
Garvie}, the authors have studied the pattern formation arising from
special initial conditions. They indicated that the spatiotemporal
dynamics of a diffusion-reaction system depends on the choice of
initial conditions. And the initial conditions are deliberately
chosen to be asymmetric in order to make any influence of the
corners of the domain more visible~\citep{medvinsky:311}. The
initial localized introduction of predators into a uniform
distribution of prey led to the spread of predators over the
domain~\citep{Garvie}. An important new feature in the
two-dimensional solutions is the way in which asymmetries in the
initial introduction of predators are reflected in the long-time
solutions.

\begin{figure*}[htp]
\includegraphics[width=14cm]{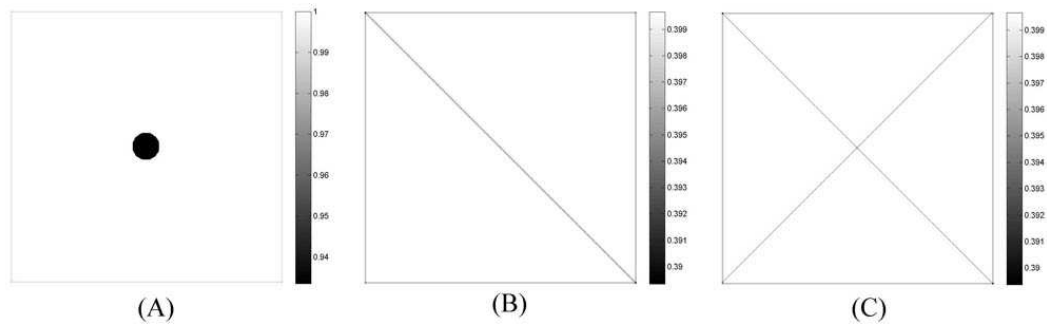}
\caption{\label{fig5} Three categories of initial perturbations
corresponding with figures 6---9.}
\end{figure*}

Based on the discussion above, we employ three categories of initial
perturbations (figure~\ref{fig5}) for further learning the evolution
of the spatial pattern of prey of model~\eqref{eq:2}. In both parts
of figure~\ref{fig5}, predators were initially introduced in a
spatially asymmetric manner. In the following, model~\eqref{eq:2}
was solved numerically with a time stepsize of $\Delta t=1$ and
space stepsize $\Delta h={1}/{3}$.

In the first, predators are introduced in a small, localized region
of the circle domain (Figure~\ref{fig5}(A)):
\begin{equation}{\label{eq:10}}
\begin{array}{l}u(x,y,0)=1.0,\\[4pt]
v(x,y,0)=\left\{\begin{array}{ll}
0.2&\qquad(x-200)^2+(y-200)^2<400,\\\noalign{\medskip}0&\qquad
{otherwise}\end{array}\right.
\end{array}
\end{equation}

The second category of initial perturbation that we have used is the
introduction of predators along a line, which is otherwise in the
steady state $(u^*, v^*)$ (figure~\ref{fig5}(B)):
\begin{equation}{\label{eq:11}}
\begin{array}{l}
u(x,y,0)=u^*+0.005\cdot\exp\Bigl(-(x-200)^2-(y-200)^2\Bigr),\\[4pt]
v(x,y,0)=v^*-0.005\cdot\exp\Bigl(-\left| x-y \right|\Bigr).
\end{array}
\end{equation}

Thirdly, we employ the so-called pitchfork initial conditions
(figure~\ref{fig5}(C)):
\begin{equation}{\label{eq:12}}
\begin{array}{l}
u(x,y,0)=u^*+0.005\cdot\exp\Bigl(-(x-200)^2-(y-200)^2\Bigr),\\[4pt]
v(x,y,0)=v^*-0.005\cdot\exp\Bigl(-\sqrt{\left|(x-y)(x+y-400)
\right|}\Bigr).
\end{array}
\end{equation}

\begin{figure}[htp]
\includegraphics[width=14cm]{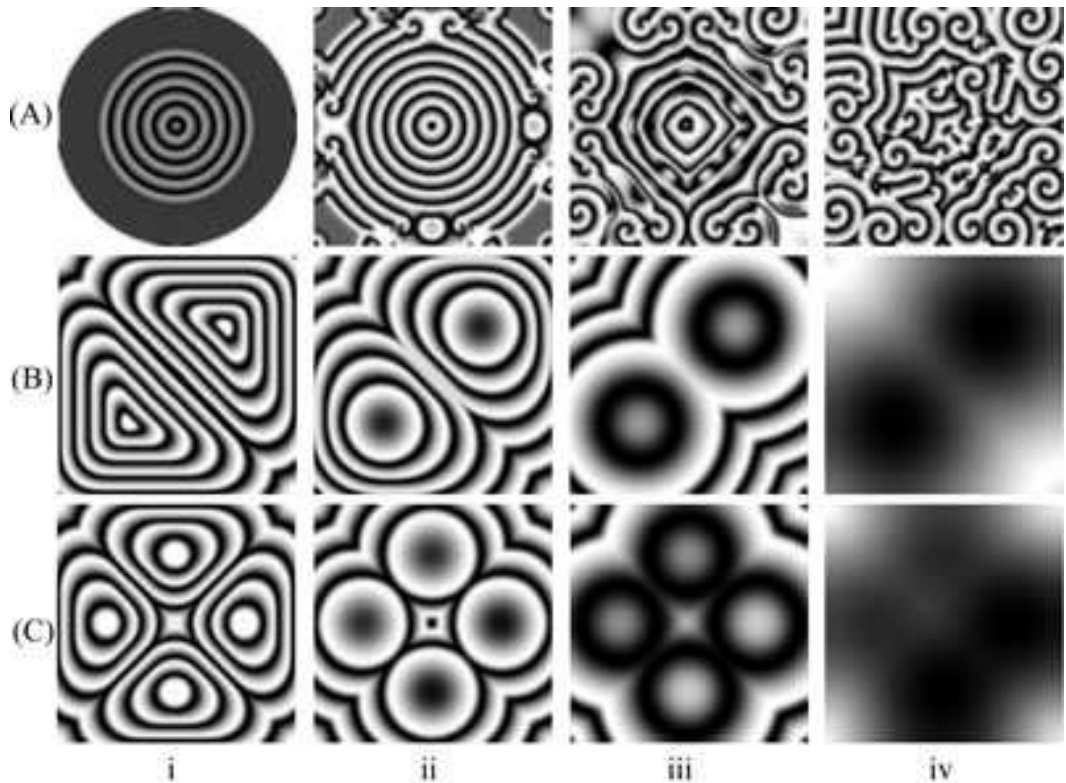}
\caption{\label{fig6}Grey-scaled snapshots of contour pictures of
the time evolution of the prey of system~\eqref{eq:2} at different
instants with $\gamma_{H}<\gamma=6.0<\gamma_{T}$. (A)(B)(C) are the
three category of initial conditions corresponding to Figure
~\eqref{fig4} and the iterations are: (i) 3000; (ii) 5000; (iii)
10000; (iv) 50000. }
\end{figure}

\begin{figure}[htp]
\includegraphics[width=14cm]{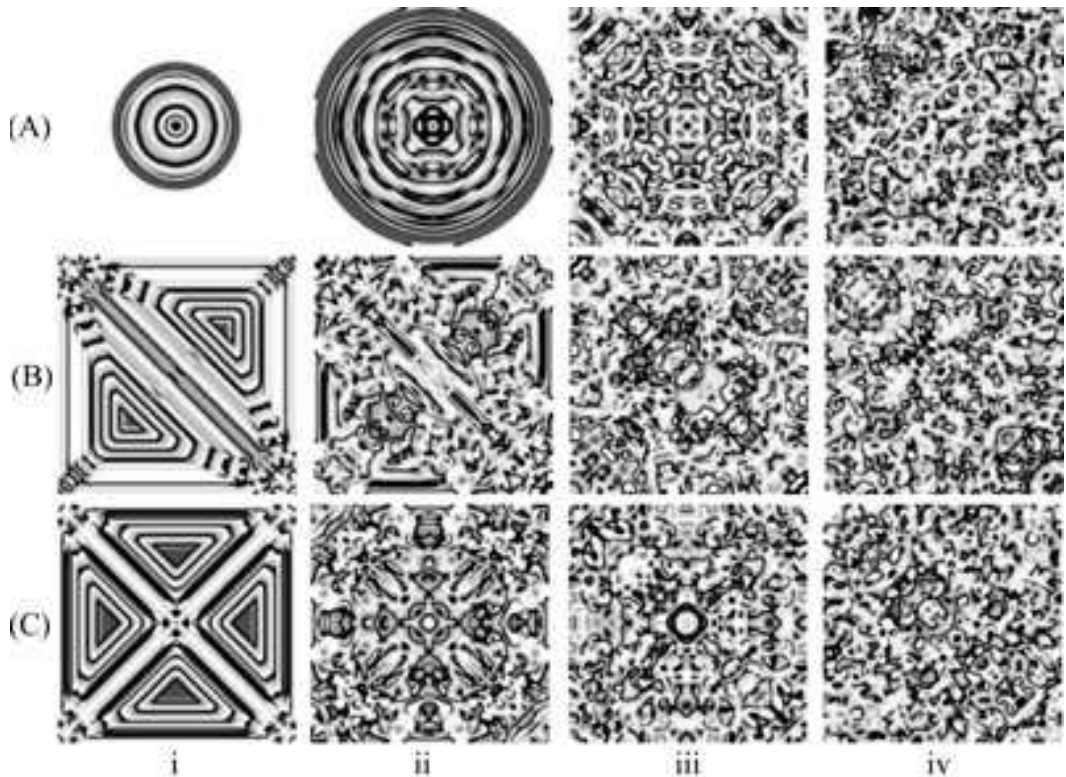}
\caption{\label{fig7}Grey-scaled snapshots of contour pictures of
the time evolution of the prey of system~\eqref{eq:2} at different
instants with $\gamma=8.0>\gamma_{T}$. (A)(B) (i) 1500; (ii) 3000;
(iii) 5000; (iv) 50000. (C) (i) 1000; (ii) 3000; (iii) 5000; (iv)
50000.}
\end{figure}

The numerical simulations results of pattern formation of model
~\eqref{eq:2} with above three categories of initial perturbations
are shown in figures~\ref{fig6} $(\gamma=6.0)$ and~\ref{fig7}
$(\gamma=8.0)$.


In figure~\ref{fig6}(A), in the case of the first category of
initial perturbation which predators are introduced in a small and
localized region of the circle domain, one can see that after a
symmetrical target pattern (figure~\ref{fig6}(A)(i)), it grows
slightly and the spiral pattern (exterior) with target pattern
(interior) emerges (figure~\ref{fig6}(A)(ii, iii)), finally with the
appearance of spiral pattern in the whole domain
(figure~\ref{fig6}(A)(iv)). And in figure~\ref{fig7}(A),
$\gamma=8.0$, with the same initial conditions, a target pattern
(figure~\ref{fig7}(A)(i)) emerges, the destruction of the target
begins from the center, and leads to the formation of the spiral
pattern (interior) with target pattern (exterior)
(figure~\ref{fig7}(A)(ii, iii)), finally, the chaotic spatial
pattern prevails the whole domain (figure~\ref{fig7}(A)(iv)).
Comparing Figure~\ref{fig6}(A)(iv) with figure~\ref{fig7}(A)(iv), we
find that Hopf instability leads to the formation of spiral patterns
and the Turing instability destroys the spiral pattern and leads to
the formation of chaotic spatial patterns. Moreover, in these two
cases, the initial nonuniformity spreading outwards through the
domain from the center provides additional evidence for
spatiotemporal pattern.

In figure~\ref{fig6}(B), with initial condition~\eqref{eq:11},
bi-target pattern (figure~\ref{fig6}(B)(i)) emerges, and finally,
the phase waves appear in the whole domain (figure~\ref{fig6}(B)
(iv)). And in figure~\ref{fig7}(B), more differently, after a target
pattern(figure~\ref{fig7}(B) (i)), a chaotic spatial pattern occurs
(figure~\ref{fig7}(B)(iii, iv)).

In the case of the third category initial
perturbation~\eqref{eq:12}, one can see that figure~\ref{fig6}(C)
and~\ref{fig7}(C) follow similar scenario to the previous case,
respectively (figures~\ref{fig6}(B) and~\ref{fig7}(B)). The
differences are that there are four-target patterns
(figure~\ref{fig6}(C)(i) and~\ref{fig6}(C)(ii)) while in the
previous case bi-target patterns occur. Comparing these two cases
with the initial perturbations defined by equations~\eqref{eq:11}
and~\eqref{eq:12}, we find that in the beginning of evolution of the
spatial pattern of prey, the special initial conditions have an
effect on the formation of spatial patterns, though the effect is
less and less with the more and more iterations.

\begin{figure}[htp]
\includegraphics[width=8cm]{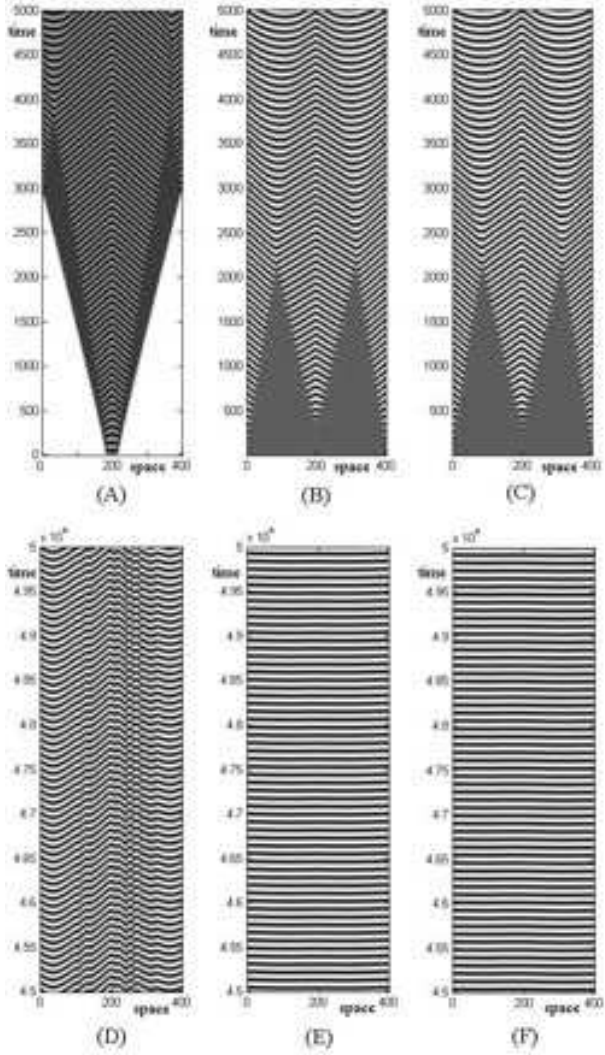}
\caption{\label{fig8} Space-time plots of variable $\gamma=6.0$.
Other parameters are the same as those in Figure\ref{fig2}
and~\ref{fig6}. }
\end{figure}

\begin{figure}[htp]
\includegraphics[width=8cm]{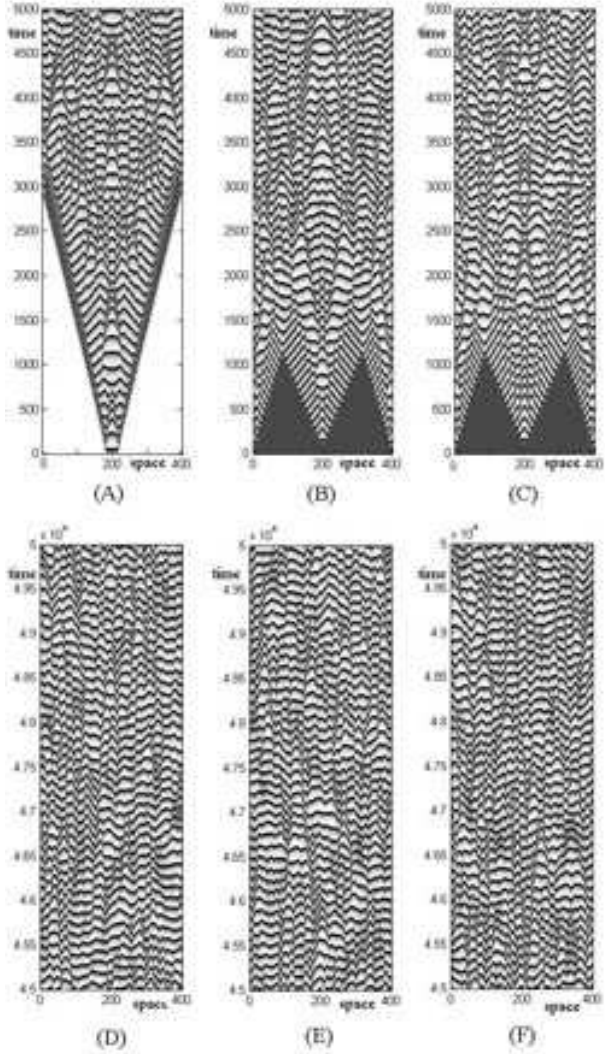}
\caption{\label{fig9}Space-time plots of variable $\gamma=8.0$.
Other parameters are the same as those in Figure~\ref{fig3}
and~\ref{fig7}.}
\end{figure}

In order to make it more clearer, we show space-time plots in
figures~\ref{fig8}($\gamma=6.0$) and~\ref{fig9}($\gamma=8.0$). The
method of space-time plots is that let $y$ be a constant (here,
$y=200$, the center line of each snapshots), from each pattern
snapshots, choose the line $y=200$, and pile these lines
in-time-order. The space-time plots show the evolution process of
the prey $u$ throughout time $t$ and space $x$. On the other hand,
we have found that the distributions of predator and prey are always
of the same type. In this section, we show the distribution of prey,
for instance.


In figure~\ref{fig8}, $\gamma=6.0$, other parameters equal
figures~\ref{fig2} and~\ref{fig6}, with the three categories of
initial perturbations (figure~\ref{fig5}), space-time plots at
different times are shown. The first row in figure~\ref{fig8}(A--C)
displays the time evolution of the prey with the iterations from 0
to 5000, while the second row (figure~\ref{fig8}(D--F)) displays the
time evolution of the prey with the iterations from 45000 to 50000.
The three columns correspond to the three categories of initial
perturbations (figure~\ref{fig5}). Form figure~\ref{fig8} and
figure~\ref{fig6}, one can see that when $\gamma=6.0$, in this case,
the Hopf instability occurs, the circular initial condition creates
spiral wave (figure~\ref{fig6}(A) and figure~\ref{fig8}(D)), and the
other two initial conditions create phase wave (figure~\ref{fig6}(B,
C) and figure~\ref{fig8}(E, F)). Furthermore, in this case, the
second and third initial conditions have the same effect on pattern
formation (figure~\ref{fig8}(E, F)).

When $\gamma=8.0$, other parameters equal figures~\ref{fig3}
and~\ref{fig7}, space-time plots at different times with the three
categories of initial perturbations (figure~\ref{fig5}) are shown in
figure~\ref{fig9}. In this case, both Hopf and Turing instability
occur. Figure~\ref{fig9}(A--C), displays the time evolution of the
prey with the iterations from 0 to 5000, while the second row
(figure~\ref{fig9}(D--F)) displays the time evolution of the prey
with the iterations from 45000 to 50000. The three columns
correspond to the three categories of initial perturbations
(figure~\ref{fig5}). Form figure~\ref{fig9} and figure~\ref{fig7},
one can see that all three initial conditions have the same effect
on pattern formation (figure~\ref{fig7}(iv) and
figure~\ref{fig9}(D--F)).

In our previous work~\citep{Wang2007}, we indicated that for the
ratio-dependent predator-prey system with Michaelis-Menten-type
functional response, the case of predator-dependent, the stationary
patterns did not dependent on the initial conditions. One can see
that, in the case of prey-dependent type~\eqref{eq:2}, pattern
formation do depend on the initial conditions. This is a major
difference between prey-dependent and predator-dependent
predator-prey models.

\section{Discussions}

In this paper, we have presented a theoretical analysis of
evolutionary processes that involve organisms distribution and their
interaction of spatially distributed population with local
diffusion. And the numerical simulations were consistent with the
predictions drawn from the bifurcation analysis, that is, Hopf
bifurcation and Turing bifurcation.

If the parameter $\gamma$ in the domain II, the Hopf instability
occurs, the destruction of the pattern begins from the interior,
while it begins from the exterior if $\gamma$ located in the domain
III, both Hopf and Turing instabilities occur. From an ecological
viewpoint~\citep{Sherratt1997}, it shows that the initial and
relatively rapid invasion of prey by predators can be followed by
two subsequent invasions.

The effort to explain the distribution of populations in terms of
the movements of individuals is an extension of one of the most
successful applications of mathematics to ecological phenomena, the
use of diffusion model to describe dispersal. The basic idea is
that, although organisms do not move randomly, the collective
behavior of large numbers of such individuals may be
indistinguishable (at the scale of the population) from what they
did~\citep{Levin1992}. In references~\citep{Levin1992,Savill1999,
David2002}, the authors indicated that the basic idea of
diffusion-driven instability in a reaction-diffusion system can be
understood in terms of an activator-inhibitor system. The
functioning of this mechanism is based on three
points~\citep{David2002}. First, a random increase of activator
species should have a positive effect on the creation rate of both
activator and inhibitor species. Second, an increment in inhibitor
species should have a negative effect on formation rate of both
species. Finally, inhibitor species must diffuse faster than
activator species. Certainly, the reaction-diffusion predator-prey
model~\eqref{eq:2}, with Ivlev functional response and predators
diffusing faster than prey, provides this mechanism.

Spirals and curves are the most fascinating clusters to emerge from
the predator-prey model. A spiral will form from a wave front when
the rabbit line (which is leading the front) overlaps the pursuing
line of predator. The prey on the extreme end of the line stop
moving as there are no predator in their immediate vicinity. However
the prey and the predator in the center of the line continue moving
forward. This forms a small trail of prey at one (or both) ends of
the front. These prey start breeding and the trailing line of prey
thickens and attracts the attention of predator at the end of the
fox line that turn towards this new source of prey. Thus a spiral
forms with predator on the inside and prey on the outside. If the
original overlap of prey occurs at both ends of the line a double
spiral will form. Spirals can also form as a prey blob collapses
after predator eat into it~\citep{Hawick2006}.

And a random increase of activator species (prey, $u$) has a
positive effect on the creation rate of both activator and inhibitor
species. Random fluctuations may cause a nonuniform prey density.
This elevated prey density has a positive effect both on prey and
predator population growth rates. From model~\eqref{eq:1}, we can
obtain per capita rates:
\begin{eqnarray}\label{eq:13}
\begin{array}{l}
 \frac{1}{u}\frac{\partial u}{\partial t}=(1-u)-\frac{v}{u}(1-e^{-\gamma u}),
 \qquad
 \frac{1}{v}\frac{\partial v}{\partial t}=\beta (\alpha-1-\alpha e^{-\gamma u}).
\end{array}
\end{eqnarray}

Since the first equation in~\eqref{eq:13} is a one-humped function
of prey density $u$, prey growth rate can be increased by a higher
local prey density at least in a range of parameter values. The
second equation in~\eqref{eq:13}, predator numerical response, is an
ever-increasing function of $u$, and high prey density always has a
positive influence on predator growth. More importantly, inhibitor
species (predator, $v$) must diffuse faster than activator species
(prey, $u$), for an increment in inhibitor species may have a
negative effect on formation rate of both species. Thus, as random
fluctuations increase local prey density over its equilibrium value,
prey population undergoes an accelerated growth. Simultaneously,
predator population also increases, but as predators diffuse faster
than prey, they disperse away from the center of prey outbreaks. If
relative diffusion ($d_2/d_1$) is large enough, prey growth rate
will reach negative values and prey population will be driven by
predators to a very low level in those regions. In other words,
where the prey density is at their maximal value diffusion will
lower the prey density at that point. Conversely, where the prey
density is at their minimal value diffusion will increase the prey
density at that point. That is, prey flow from high density to low
density regions in space. Moreover, the faster the diffusion the
greater the flow. When the prey density is high (in fact, when
$\nabla^2u<0$) proportionately less of the slower diffusing prey
leaves these points in space than the faster diffusion prey and,
therefore, the proportion of the slower diffusing prey at that point
increases more than the proportion of the faster diffusing prey.
Conversely, at low prey density (when $\nabla^2u>0$) proportionately
more faster diffusing prey enter a point in space and the proportion
of these prey increases more than the proportion of the slower
diffusing prey. Hence, at a given position in space, when the prey
density is high the proportion of slower diffusing prey increases
and when the prey density is low the proportion of the faster
diffusing prey increases. Therefore we see oscillations in the
proportions of the prey at the same frequency as the oscillations of
the density waves. The final result is the formation of patches of
high prey density is surrounded by areas of low prey density. And
predators follow the same pattern.

On the other hand, in two-dimensional reaction-diffusion systems,
rotationally symmetric patterns, known as targets or sinks, and a
generalization of them with broken circular symmetry, spirals are
being investigated experimentally as well as theoretically in many
nonlinear systems. The Belousov-Zabotinsky reaction is a well
investigated excitable reaction-diffusion system that shows all
these patterns. Spirals are characteristic patterns in slime mold
aggregates and are an important observation in cardiac arrythmias as
well. Targets and spirals, which are generally found to form around
some defects, precede some defect mediated chaos, commonly known as
spiral defect chaos~\citep{Bhattacharyay}.

Particularly, spiral patterns are being investigated theoretically
in a number of reaction-diffusion predator-prey systems, such as
Holling-type model~\citep{Savill1999,Malchow2000},
Ivlev-type~\citep{Sherratt1995,Sherratt1997,Sherratt2000,Pearce,
Garvie,Preedy2006,Uriu2007}, and so on. The functional responses of
these predator-prey models are all prey-dependent. It is necessary
or coincident?

It is well known that, for reaction-diffusion predator-prey systems,
under suitable conditions, the destabilized uniform distributions
give way to stable nonuniform patterns, which can provide the local
information of that specifies patterns of
differentiation~\citep{Levin1992}. In reference~\citep{David2002},
the authors also indicted that a simple general model for
predator-prey dynamics with predator-dependent functional response,
a reaction-diffusion system that could develop diffusion-driven
instabilities. On the contrary, if the functional response depends
only on prey density, diffusion instabilities are not possible. In
fact, in the case of predator-dependent, the interactions between
dispersing populations induce spatial heterogeneity and/or temporal
fluctuations through so-called self-structuring without help from
external forcing, and the patterns are endogenous. While the
prey-dependent predator-prey models are self-oscillation ones which
are called oscillatory systems, there are typical patterns including
spiral waves, turbulence, and target patterns. This is the reason
why we can find the spiral and target waves in model~\eqref{eq:2}.

From the theoretical study on the three-dimensional
patterns~\citep{Teemu2003, Teemu2004}, it is possible that the
three-dimensional patterns can be reflected by the two-dimensional
patterns. And two-dimensional patterns might be sufficient to
understand the general properties of dissipative
structures~\citep{Shoji}. So our two-dimensional spatial patterns
may indicate the vital role of pattern formation in the
three-dimensional spatiotemporal organization of the predator-prey
system.

Furthermore, from references~\citep{Sherratt1999,Byrne2006}, we
think that pattern formation of spatial model~\eqref{eq:2} with
special choice initial conditions~\eqref{eq:10}--\eqref{eq:12} can
be used to explain other diffusion process, such as tumor growth,
and so on.

On the other hand, the ecosystem is so complicated that we cannot
use a single method to study. We must use mixed methods, such as
analytical or experimental or numerical method.
\bigskip\bigskip\bigskip\bigskip

\bigskip\centerline{\bf Acknowledgments}\bigskip

{\small This work is by the Knowledge Innovation Project of the
Chinese Academy of Sciences (KZCX2-YW-430), the National Basic
Research Program (2006CB403207) and the Youth Science Foundation of
Shanxi Provence (20041004).}


\begin{thebibliography}{}

\bibitem[Abram \& Ginzburg(2000)]{Abram2000}
Abrams P., and Ginzburg L., 2000. The nature of predation: prey
dependent, ratio dependent or neither? Trends Ecol. Evol. 15,
337-341.
\bibitem[Alonso, Bartumeus \& Catalan(2002)]{David2002}
Alonso, D., Bartumeus, F., Catalan, J., 2002. Mutual interference
between predators can give rise to Turing spatial patterns, Ecology
83, 28-34.
\bibitem[Arditi \& Ginzburg(1989)]{Arditi1989}
Arditi R., and Ginzburg L., 1989. Coupling in predator-prey
dynamics: Ratio-dependence, J. Theor. Biol. 139, 311-326.
\bibitem[Baurmann, Gross, \& Feudel(2007)]{Baurmann}
Baurmann, M., Gross, T., Feudel, U., 2007, Instabilities in
spatially extended predator-prey systems: Spatio-temporal patterns
in the neighborhood of Turing-Hopf bifurcations, J. Theo. Biol. 245,
220-229.
\bibitem[Beddington(1975)]{Beddington}
Beddington, J., 1975. Mutual interference between parasites or
predators and its effect on searching efficiency, J. Anim. Ecol. 44,
331¨C340.
\bibitem[Ben-Jacob \& Levine(2001)]{Ben-Jacob}
Ben-Jacob, E., Levine, H., 2001. The artistry of nature, Nature,
409, 985-986.
\bibitem[Bhattacharyay(2001)]{Bhattacharyay}
Bhattacharyay, A., 2001. Spirals and targets in reaction-diffusion
systems, Phys. Rev. E. 64, 016113(4).
\bibitem[Byrne et al(2006)]{Byrne2006}
Byrne, H. M., Alarcon, T., Owen, M. R., Webb, S. D., and Maini, P.
K., 2006. Modelling aspects of cancer dynamics: a review, Phil.
Trans. R. Soc. A, 364, 1563-1578.
\bibitem[Callahan \& Knobloch(1999)]{Callahan}
Callahan, T., Knobloch, E., 1999. Pattern formation in the
three-dimensional reaction-diffusion systems, Phys. D 132, 339-362.
\bibitem[Cantrell \& Cosner(2003)]{Cantrell2003}
Cantrell, R., and Cosner, C., 2003. Spatial Ecology via
Reaction-Diffusion Equations, John Wiley \& Sons, Ltd., Chichester,
England.
\bibitem[Crowley \& Martin(1989)]{Crowley}
Crowley P., Martin E., 1989. Functional responses and interference
within and between year classes of a dragonfly population, J. North
Amer. Bent. Soc., 8, 211-221.
\bibitem[DeAngelis, Goldstein \& Neill(1975)]{DeAngelis}
DeAngelis D., Goldstein R., and Neill R., A model for trophic
interaction, Ecology 56, 881-892.
\bibitem[Garvie(2007)]{Garvie}
Garvie, M., 2007. Finite-difference schemes for reaction-diffusion
equations modelling predator-prey interactions in matlab, Bull.
Math. Biol. 69, 931-956.
\bibitem[Griffith \& Peres-Netob(2006)]{Daniel2006}
Griffith, D., Peres-Netob, P., 2006. Spatial modeling in ecology:
the flexibility of eigenfunction spatial analyses, Ecology 87,
2603-2613.
\bibitem[Hassell \& Varley(1969)]{Hassell}
Hassell M., and Varley C., 1969. New inductive population model for
insect parasites and its bearing on biological control, Nature 223,
1133-1177.
\bibitem[Hawick, James \& Scogings(2006)]{Hawick2006}
Hawick, K., James, H., Scogings, C., 2006. A zoology of emergent
life patterns in a predator-prey simulation model, Technical Note
CSTN-015 and in Proc. IASTED International Conference on Modelling,
Simulation and Optimization, September 2006, Gabarone, Botswana.
507-115.
\bibitem[Holling(1959a)]{Holling1}
Holling C., 1959. The components of predation as revealed by a study
of small mammal predation of the european pine sawfly, Cana. Ento.
91, 293-320.
\bibitem[Holling(1959b)]{Holling2}
Holling C., 1959. Some characteristics of simple types of predation
and parasitism, Cana. Ento. 91, 385-395.
\bibitem[Ivlev(1961)]{Ivlev}
Ivlev, V., 1961. Experimental ecology of the feeding fishes, Yale
University Press, New Haven.
\bibitem[Jost(1998)]{Jost1998}
Jost, C., 1998. Comparing predator-prey models qualitatively and
quantitatively with ecological time-series data, PhD-Thesis,
Institute National Agronomique, Paris-Grignon.
\bibitem[Kay \& Sherratt(2000)]{Sherratt2000}
Kay, A., Sherratt, J., 2000. Spatial noise stabilizes periodic wave
patterns in oscillatory systems on finite domains, SIAM J. Appl.
Math. 61, 1013-1041.
\bibitem[Koch \& Meinhardt(1994)]{Koch1994}
Koch, A., Meinhardt, H., 1994. Biological pattern formation: from
basic mechanisms to complex structures, Revi. Mode. Phys. 66,
1281-1507.
\bibitem[Kooij(1996)]{Kooij}
Kooij, R., 1996. A predator-prey model with Ivlev's functional
response, J. Math. Anal. Appl. 198, 473-489.
\bibitem[Kuang \& Beretta(1998)]{kuang98global}
Kuang Y., Beretta E., 1998. Global qualitative analysis of a
ratio-dependent predator-prey system, J. Math. Biol. 36, 389-406.
\bibitem[Lepp\"anen(2004)]{Teemu2004}
Lepp\"anen, T., 2004. Coputational studies of pattern formation in
turing systems, Phd-thesis, Helsinki University of Technology,
Finland.
\bibitem[Lepp\"anen et al(2003)]{Teemu2003}
Lepp\"anen, T., Karttunen, M., Kaski, K., Barrio, R., 2003.
Dimensionality effects in turing pattern formation, Inter. J. of
Mod. Phys. B 17, 5541-5553.
\bibitem[Levin(1992)]{Levin1992}
Levin, S., 1992. The Problem of Pattern and Scale in Ecology,
Ecology 73, 1943-1967.
\bibitem[Liu \& Jin(2007)]{LiuJSM}
Liu, Q.-X., Jin, Z., 2007. Formation of spatial patterns in epidemic
model with constant removal rate of the infectives, J. Stat. Mech.
P05002.
\bibitem[Malchow et al(2000)]{Malchow2000}
Malchow, H., Radtke, B., Kallache, M., Medvinsky, A., Tikhonov, D. ,
Petrovskii, S.,  2000. Spatio-temporal pattern formation in coupled
models of plankton dynamics and fish school motion. Nonl. Anal.:
Real World Appl. 1, 53-67.
\bibitem[Maini, Baker, Chuong(2006)]{Maini2006}
Maini, P., Baker, R., Chuong, C., 2006. The Turing model comes of
molecular age, Science 314, 1397-1398.
\bibitem[May(1981)]{May1981}
May, R., 1981. Stability and complexity in model ecosystems.
Princeton University Press, American.
\bibitem[Medvinsky et al(2002)]{medvinsky:311}
Medvinsky A., Petrovskii S., Tikhonova I., Malchow H., Li B-L.,
2002. Spatiotemporal complexity of plankton and fish dynamics, SIAM
Review 44, 311-370.
\bibitem[Metz \& Diekmann(1986)]{Metz1986}
Metz, J., Diekmann, O. 1986. A gentle introduction to structured
population models: three worked examples. In: The dynamics of
ph.siologicall. structured populations (Lecture Notes in
Biomathematics 68) (ed. Metz, J., Diekmann,O.), pp. 3-45. Berlin:
Springer.
\bibitem[Murray(2003)]{Murray2004}
Murray, J., 2003. Mathematical Biology II: Spatial Models and
Biomedical Applications. Springer, Berlin.
\bibitem[Neuhauser(2001)]{ClaudiaNeuhauser}
Neuhauser, C., 2001. Mathematical challenges in spatial ecology,
Noti. Amer. Math. Soc. 47, 1304-1314.
\bibitem[Pearce et al(2006)]{Pearce}
Pearce, I., Chaplain, M., Schofield, P., Anderson, A., Hubbard, S.,
2006. Modelling the spatio-temporal dynamics of multi-species
host-parasitoid interactions: Heterogeneous patterns and ecological
implications, J. Theo. Biol. 241, 876-886.
\bibitem[Preedy et al.(2006)]{Preedy2006}
Preedy, K., Schofield, P., Chaplain, M., Hubbard S., 2006. Disease
induced dynamics in host-parasitoid systems: chaos and coexistence,
J. R. Soc. Inte. doi:10.1098/rsif.2006.0184.
\bibitem[Ruan \& Xiao(2001)]{Ruan:1445}
Ruan, S., Xiao, D., 2001. Global Analysis in a Predator-Prey System
with Nonmonotonic Functional Response, SIAM J. Appl. Math. 61,
1445-1472.
\bibitem[Savill \& Hogeweg(1999)]{Savill1999}
Savill, N., Hogeweg, P., 1999. Competition and dispersal in
predator-prey waves, Theo. Popu. Biol. 56, 243-263.
\bibitem[Schnell, Grima \& Maini(2007)]{Maini2007}
Schnell, S., Grima, R., Maini, P., 2007. Multiscale Modeling in
Biology, Amer. Scie. 95, 134-142.
\bibitem[Sherratt, Lewis \& Fowler(1995)]{Sherratt1995}
Sherratt, J., Lewis, M., Fowler, A., 1995. Ecological chaos in the
wake of invasion, Proc. Nati. Acad. Sci. 92, 2524-2528.
\bibitem[Sherratt, Eagan \& Lewis(1997)]{Sherratt1997}
Sherratt, J., Eagan, B., Lewis, M., 1997. Oscillations and chaos
behind predatorcprey invasion: Mathematical artifact or ecological
reality? Phil. trans. Roy. Soc. Lond.-B 352, 21-38.
\bibitem[Sherratt, Perumpanani \& Owen(1999)]{Sherratt1999}
Sherratt, J., Perumpanani, A., Owen, M., 1999. Pattern Formation in
Cancer. In: On Growth and Form: Spatio-temporal Pattern Formation in
Biology, (editors: M.A.J. Chaplain, G.D. Singh, J.C. McLachlan),
John Wiley \& Sons Ltd.

\bibitem[Shoji, Yamada, Ueyama \& Ohta(2007)]{Shoji}
Shoji, H., Yamada, K., Ueyama, D., Ohta, T., 2007. Turing patterns
in three dimensions, Phys. Revi. E 75, 046212(13).
\bibitem[Sugie(1998)]{Sugie}
Sugie, J., 1998. Two-parameter bifurcation in a predator-prey system
of ivlev type, J. Math. Anal. \& Appl. 217, 349-371.
\bibitem[Tian(2006)]{Tian2006}
Tian, R., 2006. Toward standard parameterization in marine
biological modeling, Ecol. Model. 193, 363-386.
\bibitem[Tuing(1952)]{Turing1952}
Turing, A., 1952. The chemical basis of morphogenesis. Phil. trans.
Royal. Soc. Lond.-B 237 (B 641), 37-72.
\bibitem[Wang(2007)]{Wang}
Wang, H., Wang, W., 2007. The dynamical complexity of a Ivlev-type
prey-predator system with impulsive effect, Chaos, Soli. Frac. DOI:
10.1016/j.chaos.2007.02.008.
\bibitem[Wang, Liu \& Jin(2007)]{Wang2007}
Wang, W., Liu, Q.-X., Jin, Z., 2007. Spatiotemporal com plexity of a
ratio-dependent predator-prey system, Phys. Rev. E 75, 051913.
\bibitem[Yang et al(2002)]{yang:7259}
Yang, L., Dolnik, M., Zhabotinsky, A., Epstein, I., 2002. Pattern
formation arising from interactions between turing and wave
instabilities, J. Chem. Phys. 117, 7259-7265.
\bibitem[Uriu \& Iwasa(2007)]{Uriu2007}
Uriu K., Iwasa Y., 2007. Turing pattern formation with two kinds of
cells and a diffusive chemical, Bull. Math. Biol. DOI:
10.1007/s11538-007-9230-0.


\end{thebibliography}
\end{document}